\documentstyle[12pt]{article}
\begin{document}
\title{\vspace{-1.3cm}
{\small { THE PHYSICAL
SIGNIFICANCE OF SINGULARITIES IN THE  CHERN--SIMONS
FERMI LIQUID DESCRIPTION OF A PARTIALLY FILLED LANDAU LEVEL}}
 }

\author{ { Ady Stern$^*$ and Bertrand I. Halperin}\\
{\small{Physics Department, Harvard University, Cambridge, MA 02138}}}
\date{\vspace{-1cm}}
\maketitle

\begin{abstract}
We analyze the linear response of a half filled Landau level to
long wavelength and low frequency driving forces, using Fermi
liquid theory for composite fermions. This response is determined by
the composite fermions quasi--particle effective mass, $m^*$, and
quasi--particle Landau interaction function $f(\theta-\theta')$.
 Analyzing infra--red divergences of perturbation theory, we
get an exact expression for $m^*$, and conjecture the form of the
$f(\theta-\theta')$.
We then conclude that  in the limit of infinite cyclotron frequency,
and small ${\bf q},\omega$, the composite fermion excitation spectrum
is continuous for $0<\omega<\gamma \frac{e^2}{\epsilon h}q$, with
$\gamma$ an unknown number.
 For fractional
quantum Hall states near a half filled Landau level, we derive an exact
expression for the energy gap.
\end{abstract}
\vskip 0.2truein
The fermionic Chern--Simons theory of the half filled Landau level
maps the problem of interacting electrons in a strong magnetic field
onto that of composite fermions at zero magnetic field, interacting
with a Chern--Simons gauge field. This mapping opens the way to
a study of low energy excitations of the system by an application of Fermi
liquid theory.
In this paper we discuss such an application. Due
to space constraints, very few details are given. We refer the reader
to Ref. \cite{asbih} for further details.

Fermi liquid  theory outlines a formal method for
a calculation of the exact small wave--vector $\bf q$,
low frequency $\omega$ and zero
temperature  limit of linear response functions. In practice,
the exact calculation is usually intractable, and approximations are needed.
In section (1)
below, we  remind the reader of the formal method. Then, in
section (2), we attempt
to use it to extract information about the linear response of electrons
at, and near, a half filled Landau level.
\vskip 0.4truein
{\large\noindent {
1. Derivation of linear response function by Fermi liquid theory --
a review}}
\vskip 0.2truein

The Fermi liquid theory prescription for the calculation of linear response
functions can be put into the following steps:
\begin{itemize}
\item{Formulate the Hamiltonian of the system.}
\item{Calculate the effective mass of quasi--particles at the
Fermi level $m^*$
and the quasi--particle Landau
interaction function $f({\bf k}-{\bf k'})$.
This calculation is done in two stages. First, calculate
the single particle self energy $\Sigma({\bf k},\omega)$
for states near the Fermi level, and the two--particle
interaction operator $\Gamma({\bf k}_1,\omega_1;{\bf k}_2,\omega_2;
{\bf k}_1+{\bf q}, \omega_1+\Omega;{\bf k}_2-{\bf q}, \omega_2-\Omega)$
for an exchange scattering near the Fermi level. Then, use
  the following formulae,
\begin{equation}
\begin{array}{ll}
m^*=m\frac{1\ +\ \ \frac{\partial\Sigma}{\partial\omega}}
           {1-{m\over k_F}\frac{\partial\Sigma}{\partial |k|}}\\ \\
f({\bf k}-{\bf k}')=z^2\lim_{\Omega\rightarrow 0}
\Gamma({\bf k},0;{\bf k'},0;
{\bf k'},\Omega;{\bf k}, -\Omega)
\end{array}
\label{formulae}
\end{equation}
to extract $m^*$ and $f$ out of $\Sigma$ and $\Gamma$. In Eq. (\ref{formulae}),
$z$ is the single particle Green function residue at the Fermi level.

In most practical cases, the self energy and the interaction
operator cannot be calculated exactly, and have to be approximated by
a perturbative calculation.
}
\item{Consider now the response of a 2D fermion system to a driving force
in the limit ${\bf q},\omega\rightarrow 0$.
In linear response to such a  driving force, the shape of the Fermi surface
acquires a space and time dependence characterized by the same
Fourier components ${\bf q},\omega$, and  is described by a function
$\nu_{{\bf q},\omega}(\theta)$. The angle $\theta$ parametrizes
the Fermi surface, whose ground state shape is a circle.
The function
$\nu_{{\bf q},\omega}(\theta)$ describes   the distortion of the Fermi surface
relative to its  ground state shape.
 It is related to the driving force acting on the
fermions by means of
 the Boltzmann equation, in which the effective mass and
the Landau function are parameters:

\begin{equation}
(\omega-{ q}{\frac{k_F}{m^*}}\cos{\theta})\nu_{{\bf q},\omega}
(\theta)-\frac{1}{(2\pi)^2}{ q}k_F\cos{\theta}
\int d\theta' f(\theta-\theta')\nu_{{\bf q},\omega}(\theta')=
{\bf\hat k}\cdot {\bf F}_{{\bf q},\omega}
\label{eomnu}
\end{equation}
where
the angle $\theta$ is measured relative to $\bf q$;
the function
$f(\theta-\theta')$ is the quasi--particle interaction function $f({\bf k,k'})$
for two wave-vectors at the Fermi surface, at angles $\theta,\theta'$ relative
to $\bf q$;  and
$\bf F$ is the driving force. In the case of long range interaction
the driving force $\bf F$ deserves a further
discussion, which is given below.

}
\item{
In the absence of driving force (${\bf F}=0$), the solutions to the
Boltzmann equation (\ref{eomnu}) are low energy excitation modes of
the Fermi liquid. As long as $f(\theta-\theta')$ is regular, these modes
are composed of a continuum, at $0<\omega<v_F^*q$, and possibly a number
of discrete modes.
}
\item{
The
Boltzmann equation relates the driving force to the shape of the Fermi surface.
The shape of the Fermi surface determines the three--vector current
 $J_\mu({\bf q},\omega)$, according to:
\begin{equation}
\begin{array}{ll}
J_0({{\bf q},\omega})=k_F\int d\theta \nu_{{\bf q},\omega}(\theta) \\ \\
{\bf J}({{\bf q},\omega})=\int d^2k \delta(k-k_F)
{\bf u}_{\bf k}
\nu_{{\bf q},\omega}(\frac{\bf k\cdot q}{kq})
\end{array}
\label{nuj}
\end{equation}
with ${\bf u_k}\equiv
\frac{\bf k}{m^*}+\int d^2k' f({\bf k},{\bf k'})\delta(\epsilon_{\bf k'}-\mu)
\frac{\bf k'}{m^*}$.
Combined together, Eqs. (\ref{eomnu}) and (\ref{nuj}) relate the current
to the driving force, i.e., lead to the linear response functions.
}
\item{In the case of long range interactions between the fermions, the driving
force appearing in the right hand side of the Boltzmann equation is the total
driving force, composed of an externally imposed part and an internally
induced part. The induced part is exerted by charge density and current
induced by the driving force in the system.
}
\end{itemize}
\vskip 0.1truein

{\large\noindent 2.  Application to the $\nu=1/2$ problem}
\vskip 0.2truein

{\noindent\bf $\bullet$ The fermions' Hamiltonian}  is:
\begin{equation}
\begin{array}{ll}
H={1\over {2m}}
\int d^2r\Psi^+({\bf r})[-i{\bf\nabla}
-&{\bf a}({\bf r})]^2\Psi({\bf r})\\ \\
+&{1\over 2}
\int d^2r\int d^2r'[\Psi^+({\bf r})\Psi({\bf r})-{n}]V({\bf r}-{\bf r'})
[\Psi^+({\bf r'})\Psi({\bf r'})-{n}]
\\ \\
\end{array}
\label{ham}
\end{equation}
supplemented by the gauge condition ${\bf \nabla}\cdot{\bf a}=0 $
and  the constraint  ${\bf\nabla}\times{\bf a}({\bf r})=
2\pi\tilde{\phi}
[\Psi^+({\bf r})\Psi({\bf r})\\ - n ]$, where ${n}$ is the average
electron density. The parameter $\tilde\phi$ is the number of flux quanta
attached to every fermion. While $\tilde{\phi}=2$ for $\nu=1/2$, it is useful
to keep $\tilde\phi$ as a variable parameter (see below). In Eq. (\ref{ham})
we use a system of units where $\hbar={e\over c}=1$.
\bigskip

{\noindent$\bullet\ $\bf  Perturbative calculation:}
There are two dimensionless parameter in the present problem. The first,
$\tilde\phi$, the number of flux quanta attached to each fermion, is
the dimensionless fermion--gauge field coupling constant. The second is
the ratio of Coulomb and cyclotron energies, $\frac{e^2k_F}{\omega_c}$.
In the physical system $\tilde{\phi}=2$
and $\frac{e^2k_F}{\omega_c}$ is small.

We define the unperturbed problem as that in which both parameters are
zero, and the problem is that of
non--interacting fermions at zero magnetic field. In our perturbative scheme,
$\tilde\phi$ is turned adiabatically on, and so is the external magnetic field
$B$,
such that $B=2\pi\tilde{\phi}n$. In parallel, $\frac{e^2k_F}{\omega_c}$
is turned on from zero, too.
\bigskip

{\bf\noindent$\bullet$ The effective mass:}
 A calculation of the effective mass using
a low order perturbation theory approximation for the self energy leads to the
conclusion that the effective mass of a quasi--particle of energy $\omega$
above the Fermi energy diverges logarithmically for small $\omega$
\cite{Hlr}. We
have studied the effect of higher order processes on this result, and
concluded that {\it to all orders in $\tilde\phi$},
\begin{equation}
\lim_{\omega\rightarrow 0}m^*(\omega)=\frac{\tilde{\phi}^2}{2\pi}
\frac{\epsilon k_F}{e^2 }|\ln{\omega}|
\label{meff}
\end{equation}
This conclusion is based on two observations: first, following from general
principles, the infra--red limit of the gauge field propagator can be
calculated exactly. Second, by use of Ward identities, we are able
to show that the renormalization of the vertices by high--momentum exchange
processes does not affect the effective mass.
\bigskip

{\bf\noindent$\bullet$ The Landau function:}
 A calculation of the Landau function
within low order perturbation theory leads to the conclusion that
$f(\theta-\theta')=\frac{(2\pi)^2}{m}\delta(\theta)$. When substituted into
the Boltzmann equation (\ref{eomnu}), this singular form of the
Landau function leads to a re-renormalization of the effective mass from
infinity back to its bare value, and to a cancelation of the logarithmically
divergent effective mass from the linear response functions.

This cancelation is consistent with calculations of the response functions
using perturbation theory, renormalization group and bosonization \cite{Kim}.
However,
the re -- renormalization of the mass back to its {\it bare} value is
inconsistent with the requirement that the low energy spectrum of a
partially filled Landau level should be independent of the bare mass
at the limit of infinite cyclotron energy, and should depend on
the Coulomb energy scale only.
 To reconcile both
arguments, we conjecture that the correct form of the Landau function is
\begin{equation}
f(\theta-\theta')= A\delta(\theta-\theta')+{\rm regular\ \ part}
\label{truef}
\end{equation}
where $A$ is of the order of ${e^2}\over{\epsilon k_F}$.
In the limit of infinite
cyclotron energy, we expect the regular part to give a contribution of
the order of $1\over m$ to the
first two Fourier components, $f_0$ and $f_1$, only.
\bigskip

{\bf \noindent $\bullet$ Low energy linear response:}
The Boltzmann equation can be used to calculate the density--density
component of the fermions' linear response
function, $\Pi_{\mu\nu}
({\bf q},\omega)$, which gives the response to the total scalar and vector
potentials. The total potentials
 include both  the Coulomb and Chern--Simons potentials resulting
from the induced density and current fluctuations of wave vector $\bf q$
and frequency $\omega$. Poles in this polarizability function can
be loosely interpreted as excitation modes of the fermion
system, when the long range forces are suppressed.
 If our conjecture for the
Landau function (Eq. (\ref{truef})) is indeed correct, then the spectrum of
density excitations is composed mostly of a continuum of modes at $\omega<
\tilde{v_F}q$, where $\tilde{v_F}\propto\frac{e^2}{\epsilon h}$, with the
proportionality constant being an unknown number.
This continuum leads to a finite longitudinal electronic resistance
in that range of ${\bf q},\omega$. On top of that
continuum, our conjecture predicts, at the limit of infinite cyclotron
energy, 1 or 2 discrete poles, whose energy depends on the bare
mass of the electron. While these poles are not excitation modes of the
electronic system, they do have an effect on its linear response.
\bigskip

\vskip 0.25truein
{\large\noindent 3. Implications to FQHE states near $\nu=1/2$}
\vskip 0.15truein
We have studied the energy gap corresponding to a fractional quantum Hall
state of the form $\nu=\frac{p}{2p+1}$, by mapping the state onto that
of $p$ filled composite--fermion Landau levels, and directly calculating
the discontinuity in the chemical potential when the density is varied.
Similar to the case of the effective mass, we were able to obtain an
exact result, namely,
\begin{equation}
\lim_{p\rightarrow\infty}E_g=\frac{\pi }
{2}\frac{e^2}{\epsilon l_H}\frac{1}
{(2
p+1)\ln{(2p+1)}}
\label{gapp}
\end{equation}
\vskip 0.2truein
{This work was supported in part by NSF grant DMR
94--16910.
AS acknowledges financial support of the Harvard Society of Fellows.}

\end{document}